# Magnetic Field Dependence of CDW Phases in Per$_2$M(mnt)$_2$ (M=Pt, Au)

J. S. Brooks[a*], D. Graf[a], E. S. Choi[a], M. Almeida[b], J.C. Dias[b], R.T. Henriques[b], and M. Matos[c]

[a]NHMFL/Physics, Florida State University, Tallahassee, FL 32310, USA
[b]Dept. de Química, Instituto Tecnológico e Nuclear/ CFMCUL , P-2686-953 Sacavém, Portugal
[c]Dept. de Engenharia Química, Instituto Superior de Engenharia de Lisboa, P-1900 Lisboa, Portugal

Recently the authors discovered that the suppression of the charge density wave (CDW) ground states by high magnetic fields in the organic conductor series Per$_2$M(mnt)$_2$ is followed by additional high field, CDW-like phases. The purpose of this presentation is to review these compounds, to consider the relevant parameters of the materials that describe the manner in which the CDW ground state may undergo new field induced changes above the Pauli limit.

## Introduction and Background

The metal dithiolate-perylene complexes (Per)$_2$M(mnt)$_2$ (or: (C$_{20}$H$_{12}$)$_2$[M(S$_2$C$_2$(CN)$_2$)$_2$]) where Per is perylene, M represents a transition metal such as Au, Pt, Pd, Ni, Cu, Co, Fe, etc., and mnt is maleonitriledithiolate, have had an intriguing history since the report of high electrical conductivity by Alcácer and Maki in 1974[1]. More recently, a thorough review of these materials has been provided by Almeida and Henriques[2]. The crystal structure of the α-Per$_2$M(mnt)$_2$ system is shown in Fig. 1, and representative parameters for two compounds with diamagnetic (M= Au, Cu) and paramagnetic (M= Pt, Pd) anions are given in Table 1 (at the end of this paper).

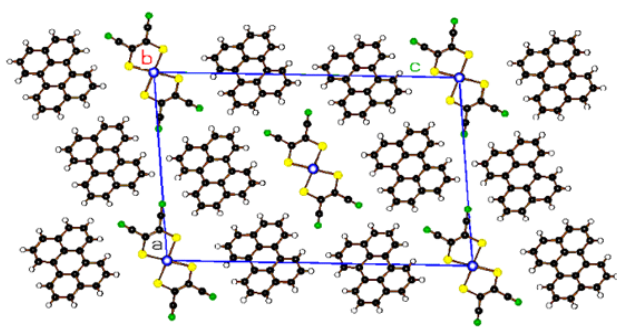

Figure 1. Crystallographic projection of (Per)$_2$M(mnt)$_2$ along the perylene *b*-axis stacking direction (After Ref. [2].)

The transfer integrals associated with the perylene chains for the a, b, and c directions are approximately 2, 150, and 0 meV respectively [3]. The conductivity (~ 700 $\Omega^{-1}$cm$^{-1}$ at room temperature) is primarily along the perylene chain stacking direction b, where the anisotropy of the conductivity $\sigma_b/\sigma_a$ is of order 10$^3$. Hence these materials are in the limit of highly one-dimensional conductors. In comparison with the Bechgaard salts[4], the interchain coupling is an order of magnitude weaker along the second and third directions, and orbital coupling due to imperfect nesting mechanisms should be unlikely. A unique aspect of the structure is the segregated two-chain nature of the compound which involves regular stacks of both the donor(cation) Per and acceptor (anion) M(mnt)$_2$ planar molecules. The 2:1 stoichiometry and the monoanionic nature of the M(mnt)$_2$ counteranions make the regular perylene stack originating a ¾ filled conducting band, and the M(mnt)$_2$ stack is considered to be insulating. Thermopower measurements are consistent with this filling. For all metals listed above, the materials undergo a transition at T$_{MI}$ to an insulating state, implying the perylene chains undergo a Peierls transition.

It cannot be over emphasized that sample quality plays an important role in determining the intrinsic properties in these materials, and that in their synthesis it is difficult to produce good quality crystals. Hence it is usually necessary to measure many samples to find those that give consistent results. This can be generally determined by the sharpness of the MI transition, which in M= Au for instance, took a significant period of time to resolve.

Peierls and Spin-Peierls Transitions: For those anions with spin (e.g. Pt = 1/2, Pd = 1/2) spin-Peierls behavior is observed in the magnetic properties (susceptibility, NMR, ESR). In the materials with paramagnetic anions, a spin-Peierls transition in the $M(mnt)_2$ chains also occurs at $T_{MI}$. Diffuse x-ray scattering[5] indicates that there is a $2k_F$ distortion (tetramerization) in the perylene chains ($2k_F = \pi/2b = b^*/4$ for 1/4 filling) that is responsible for the Peierls type MI transition at a temperature ($T_{MI}$). However, the $b^*/4$ reflections are only seen below $T_{MI}$ and are generally very weak. With the diamagnetic anions, $b^*/4$ reflections have only been seen for M = Co and Cu, perhaps due to their higher MI transition temperatures. In contrast, with the paramagnetic anions M =Pt, Pd, Ni, and Fe, a very strong $b^*/2$ ($4k_F$) reflection is observed, which generally shows precursor effects well above $T_{MI}$. The susceptibility associated with the $b^*/2$ fluctuations diverges at $T_{MI}$, where the magnetic measurements indicate a transition to a non-magnetic, spin Peierls ground state in the magnetic chains. In some cases, both $b^*/2$ and $b^*/4$ reflections below $T_{MI}$ are observed[5]. Alloy studies in the series $(Per)_2(Pt_xAu_{1-x})(mnt)_2$ indicate that the MI transition is driven by the (dominant) Peierls instability, which in turn triggers the spin-Peierls transition. It is still not clear, however, how a $2k_F$ instability involving phonons in the perylene chain can trigger a $4k_F$ instability involving spins in the magnetic chains. Some recent work involving an RKKY mechanism between the conduction electrons and the localized spins has suggested a way around this issue[6].

Evidence for a CDW Ground State: Diffuse x-ray satellites associated with the $2k_F$ distortion in the perylene chains has yet been observed in the M = Au and Pt systems, but the similarity of the physical properties with the other compounds with higher $T_{MI}$'s indicate that the Peierls transition is present. The satellites may be too weak to measure due to the low temperatures needed, sample size, exposure times, etc. Although the $2k_F$ distortion is indicative of a charge density wave (CDW) ground state, the first assignment of a CDW was made for the case of $(Per)_2Au(mnt)_2$, based on non-linear transport measurements by Lopes et al.[7]. This, and subsequent studies showed that all characteristics of CDW dynamics, including the threshold electric field for depinning, and both broad and narrow band noise were observed for M = Au, and Pt, compounds[8]. Notably, the threshold field at ~4 K below $T_{MI}$ for M = Au (~ 1V/cm or less) was significantly lower than for M = Pt (~ 10 V/cm or more). Here the spin-Peierls dimerization was proposed as a way to enhance the commensurability, and hence the pinning potential[9].

Magnetic field dependence of the Peierls ground state: The magnetic field suppression of the Peierls transition temperature in the $(Per)_2M(mnt)_2$ compounds was first reported by Bonfait et al. [10] to 8 T. Subsequent work [11] to 18 T showed that the field dependence approximately followed a mean field theory[12] description $(T_{MI}(B)-T_{MI}(0))/T_{MI}(0) = -\gamma/4(\mu B/kT_{MI}(0))^2$, where the coefficient $\gamma$ was found to be about a factor of two lower that the predicted value $\gamma \sim 1$. This difference may be explained in terms of 1D fluctuations, which can suppress $T_{MI}$ below the mean field transition temperature $T_{MF}$[13], which in this case is about a factor of two. Up to 18 T, $T_{MI}$ is only reduced by about 20%.

Summary of previous work: The $(Per)_2M(mnt)_2$ systems show a Peierls-type transition on the perylene chains at $T_{MI}$ into a charge density wave (CDW) ground state. Diffuse x-ray scattering, nonlinear transport, and the field dependence of $T_{MI}$ are all consistent with this description. When $M(mnt)_2$ is paramagnetic, an additional spin-Peierls transition occurs simultaneously on the anion chains. Although the coupling between the spin and the conducting chains is not yet understood, it is generally believed that spin-Peierls transition is induced by the lattice distortion of the Peierls transition, and that the spin system only modifies the physical properties of the CDW ground state, as seen for instance in the anisotropic field dependence of $T_{MI}$ in the M= Pt system.

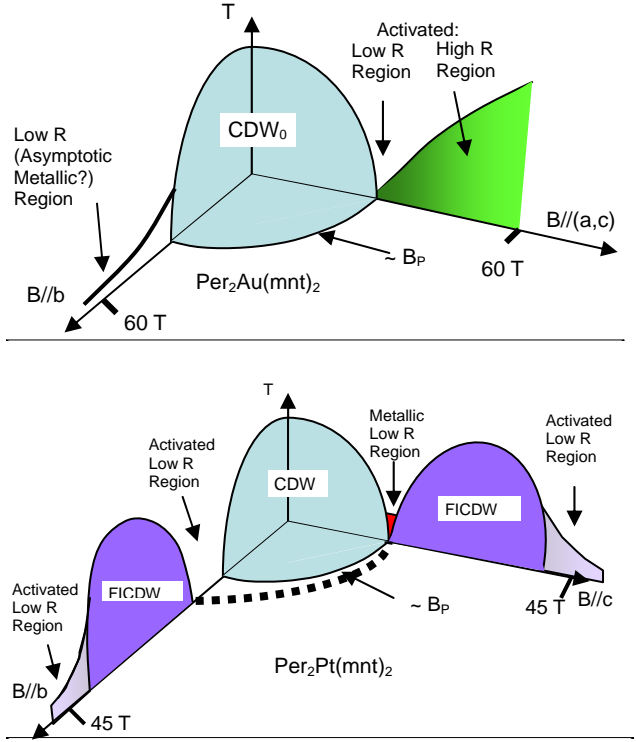

Figure 2. High field behavior of the CDW ground states in (Per)$_2$Au(mnt)$_2$ and (Per)$_2$Pt(mnt)$_2$ for B//c and B//b based on Refs. [14-16] and the present work (see text for discussion).

**High magnetic field dependence of the CDW ground states**

Theoretical Expectations: Although the mean field theory[12] gives a reasonable description of the suppression of $T_{MI}$ at lower fields, new theoretical descriptions are necessary near and above the Pauli limit. Fujita et al. have shown that in the absence of interchain bandwidth, where no orbital effects enter, a commensurate CDW can undergo a transition to an incommensurate CDW (ICDW) in the presence of a high magnetic field[17]. Here a soliton-like structure can arise, and the ICDW gap will generally decrease with increasing magnetic field. An increase in the magnetization at the CDW-ICDW transition is also predicted[17]. Zanchi, Bjelis, and Montambaux [18] treated a CDW ground state with a highly anisotropic Hubbard model where both spin density wave (SDW) and CDW interactions were included. An essential feature of the model was the prediction that the ambient ground state $CDW_0$ would evolve into a high field $CDW_X$ state with increasing magnetic field. Moreover, they predicted that as the interchain bandwidth increased from a perfectly nested to a more imperfectly nested state, dramatic differences would arise in the $T_c(B)$ phase diagram. This includes a cascade of transitions when the imperfect nesting parameter is near unity. Indeed, a remarkable comparison between this theory, including the pressure dependence, has been made with the behavior of the CDW ground state of $\alpha$-(BEDT-TTF)$_2$KHg(SCN)$_4$ [19-23]. Most recently, Lebed has treated the theory of a field induced CDW (FICDW) from a metallic, Q1D ground state where, for instance, the CDW ground state is first removed by pressure [24]. A cascade of FICDW transitions are predicted, with a transition temperature lower than for the ambient CDW. Commensurate effects for tilted magnetic fields, which change the transition temperature and the frequency of the cascade, are also predicted. A general feature of all theoretical work is that in magnetic fields, the nesting vector will be modified by a Zeeman term, and for finite interchain coupling, also an orbital term.

Experimental findings: In high magnetic field electrical transport, magnetization, thermopower, and Hall effect studies, Graf and co-workers [14-16] have shown that the low temperature, high resistance CDW phases in Per$_2$M(mnt)$_2$ (for M = Au, Pt) are not only suppressed in high magnetic fields of order of the Pauli field $B_P$, but that new high field behavior appears above $B_P$. Sketches of the magnetic field dependent phase diagrams based on Refs. [21-23] and the present work are given in Fig. 2. For fields above $B_P$, the resistance rises again for both the Pt and Au compounds (except for B//b for M= Au). We have described this second, high resistance phase as a field induced charge density wave (FICDW) phase, based on comparisons with current theoretical work in this area[18, 24]. At even higher fields, the FICDW for M = Pt is re-entrant to a second low resistance state, but for M= Au the second high resistance state continues to the highest fields we have accessed (60 T). An intervening low resistance state in the vicinity of $B_P$ appears between the CDW and FICDW phases, which for some samples and field orientations appears to be activated. However, in the case of B//c for M=Pt, metallic behavior in the temperature dependent resistance has been observed below 1.8 K in the vicinity of $B_P$.

In general, theoretical expectations for the field dependence of an ambient CDW are that at higher fields, the field dependence of $T_{MI}$ will deviate from mean-field predictions, and near or above the Pauli limit a modified CDW state can appear where $T_{MI}$ will nevertheless continue to decrease at higher fields. Moreover, due to the highly 1D nature of the $Per_2M(mnt)_2$ compounds, effects due to orbital anisotropy should be minimal, and primarily driven by an isotropic Zeeman mechanism. In light of this, the finding of Graf et al. that new high resistance, high field phases appear in both compounds that are sensitive to magnetic field direction, is surprising. Referring first to the $Per_2Au(mnt)_2$ system, we find that $T_{CDW}$ decreases with increasing field in agreement with both mean field and Hubbard-type calculations. At, and above the Pauli field $B_P$, the resistance is still thermally activated, although there are dramatic differences in the activation energy with field orientation[14] above $B_P$. Recent pulsed field measurements to 60 T [16] suggest that for B//b (i.e. the field is along the chain direction), the ground state may be metallic in the high field limit, and further experiments are planed to check this possibility. However, when there is a component of the magnetic field perpendicular to the chain direction, a second, high resistance state appears at higher fields with an orientation dependent activation energy[14], which even at 60 T is still increasing in resistance. High pressure shifts this behavior to lower fields[16]. The M = Au system gave us the first indication that orbital effects may be present in these compounds, even though they are expected to be nearly perfectly one dimensional.

Turning to $(Per)_2Pt(mnt)_2$, the suppression of the $CDW_0$ ground state in high magnetic fields is similar to that in the M=Au compound, but several very important differences appear above $B_P$. The most pronounced difference is that the second, high resistance state (FICDW) that appears above $B_P$ does not continue to increase, but is re-entrant to a lower resistance state in the high field limit. Moreover, unlike the case for M = Au, the FICDW appears for all field orientations, including B//b. Arrhenius analysis from the data of Ref. [15] of the temperature dependent resistance in the FICDW yields activation energies of 12 K, 30 K, and 15 K for fields at $B_P$, at the center of the FICDW, and in the high field re-entrant limit, respectively. Hence the magnetic field induces changes in the FICDW gap that are maximum in the center of the FICDW phase. Preliminary thermopower and Hall effect data for M=Pt also show structure corresponding to the main features in the FISDW range[16]. Another phenomena that has only been observed in the M=Pt material, and only for B//c, is that the resistance in the vicinity of $B_P \sim 23.5$ T drops to a very low value, with a metallic temperature dependence in the low temperature limit[16]. This has only been observed for B//c, i.e. for the field perpendicular to the most conducting plane.

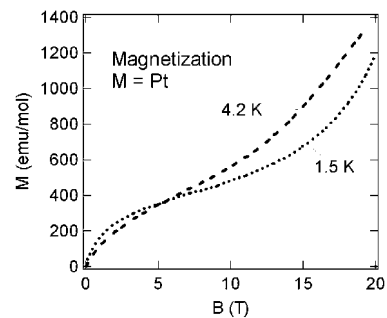

Figure 3. Magnetization of $(Per)_2Pt(mnt)_2$ at 1.5 and 4.2 K. (After Ref.[25] )

Torque Magnetization: For the magnetization, and for the M = Pt samples in particular, the spin-Peierls ground state should be sensitive to high fields[6]. Magnetization has been carried out to 20 T at low temperatures on a polycrystalline sample by Henriques[25] as shown in Fig. 3. Although the data has a sigmoid shape, there is no pronounced feature associated with either spin-flop or spin-flip processes. At the lowest temperature the sigmoid pattern does however sharpen at both low and high fields. In our measurements, an AFM cantilever is used to sense the torque magnetization on single crystals[26]. Although sensitive to changes in magnetization, this method cannot easily provide an absolute calibration, and the sensitivity depends on both the field intensity, and the position of the cantilever with respect to the field direction. The magnetization is related to the ratio torque/B which becomes highly uncertain for $B \approx 0$. Torque data for M = Pt is shown in Fig. 4 to 33 T for a sample rotation from b into the a-c plane. The overall behavior involves some apparent structure below 1 T, followed by a monotonic increase in signal, which above 15 T begins increase more rapidly. In the range 20 T and above peaks appear in the torque signal, labeled $P_1$ and $P_2$, and the field positions of these peaks vs. θ are given in the inset. The sign reversal of the signal is due to the plane of the cantilever with respect to magnetic field. As shown in Fig. 5, a small hysteretic effect is observed in the high field region which relaxes to a central value over a period of about 2 seconds (not associated with the time constants of the electronics).

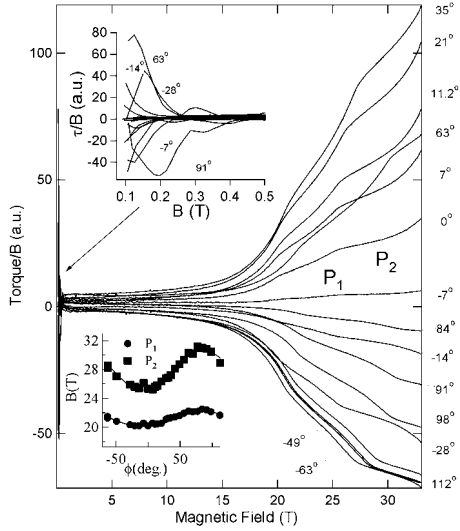

Figure 4. Torque magnetization ($\tau/B$) of $(Per)_2Pt(mnt)_2$ vs. angle at 0.5 K. Note: $0^0 \sim B//(a-c)$; $90^0 \sim B//b$. Upper inset: detail of low field signal. Lower inset: angular dependence of $P_1$ and $P_2$ features.

In Fig. 6 the simultaneous measurement of two samples, one by magnetization and the other by transport, are shown for two field directions to 45 T, where an additional peak $P_3$ is seen in the magnetization. The angular anisotropy of the magnetization peaks follows the anisotropy of the corresponding features in the transport. Although there is some uncertainty, the assignment (following the notation of Ref.[15]) is consistent with $P_1 \sim B_0$ (corresponding to the suppression of the CDW near the Pauli limit), $P_2 \sim B_x$ (corresponding to the onset of the FICDW), and $P_3 \sim B_e$ (corresponding to the re-entrant FICDW-High Field State). The temperature dependence of the $P_1$ and $P_2$ structures are shown in Fig. 7 where the low temperature data has been subtracted from a monotonic background term obtained by fitting the data at 4 K to a low order polynomial. The amplitude of these structures increases approximately linearly with decreasing temperature, and their positions follow the general pattern of the T-B phase diagram for the CDW and FICDW ground states[15], as shown in the inset and upper panel respectively. The main importance of the structure in the magnetization is that its correspondence with the transport studies used to map the CDW and FICDW boundaries shows that the general description of the phases shown in Fig. 2 for M = Pt are thermodynamic, and not due to artifacts associated with resistance measurements.

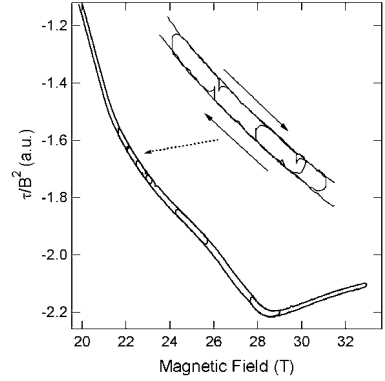

Figure 5. Detail of torque magnetization hysteresis for θ = $-63^0$ (from Fig. 4).

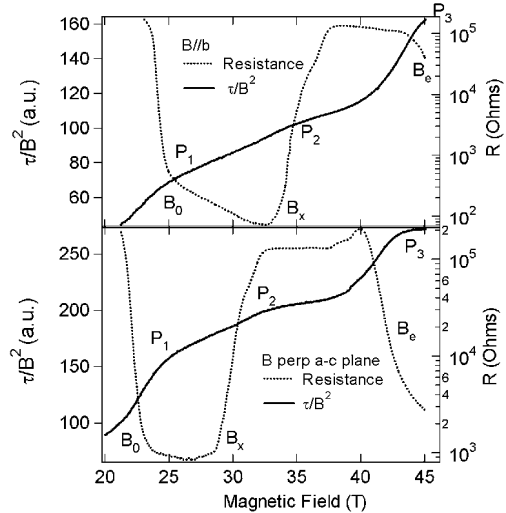

Figure 6. Correspondence of field-induced transitions in transport and torque susceptibility ($\tau/B^2$) for $(Per)_2Pt(mnt)_2$ at 0.5 K. Measurements were simultaneous, but on different samples.

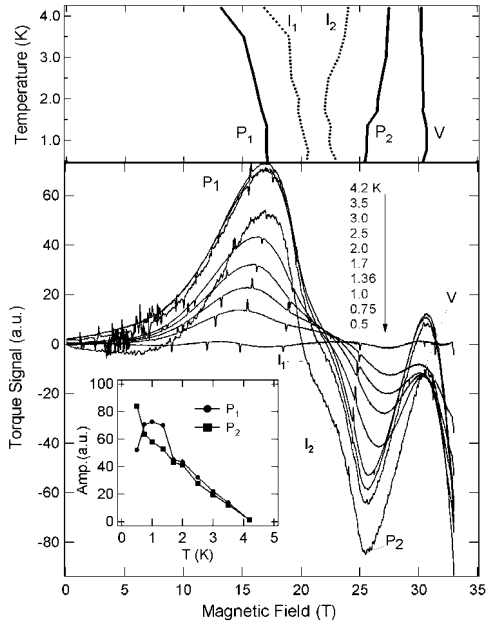

Figure 7. Temperature dependence of field-induced transitions in torque magnetization signal for $(Per)_2Pt(mnt)_2$. Inset: Temperature dependence of the absolute magnitude of the main peak features. Upper panel: Temperature dependence of the field positions of the peak (P), inflection (I), and valley (V) features.

Pressure Dependence: Motivated by the theoretical predictions of Zanchi et al.[18], where a decrease in the perfect nesting parameter decreases the $T_{MI}$ transition temperature, and also produces cascade effects at lower fields, we have begun a study of the pressure dependence of the CDW ground state in magnetic fields. The temperature dependence of the resistance for increasing pressure at zero field is similar to that previously measured by Yamaya et al.[27] in the sense that beyond about 5 kbar, the CDW appears to be gone, but there is still a non-metallic resistance at low temperatures. Using the simple scaling relationship between conductivity and bandwidth $\sigma \sim t^2$, we estimate that the average interchain bandwidth increases by about 5% at our highest pressure. Preliminary magnetoresistance (MR) studies to 18 T for both M = Pt and Au are described in Ref. [16]. Here the MR is generally positive, and exhibits some kind of structure. At this time it is not clear what the structure is, and more systematic pressure studies between 0 and 10 kbar are underway. At lower pressures it is possibly related to the cascade effects predicted for larger imperfect nesting[18], as observed in $\alpha$-$(BEDT-TTF)_2KHg(SCN)_4$[19-23], or they could possibly be quantum oscillations arising from very small pressure induced pockets in the Q1D Fermi surface. In Fig. 8 we show the magnetoresistance to 18 T for the M=Pt compound vs. rotation in field, where it is clear that the structure is following some kind of orbital behavior. Plotting the data vs. inverse field perpendicular to the a-b plane further tests this.

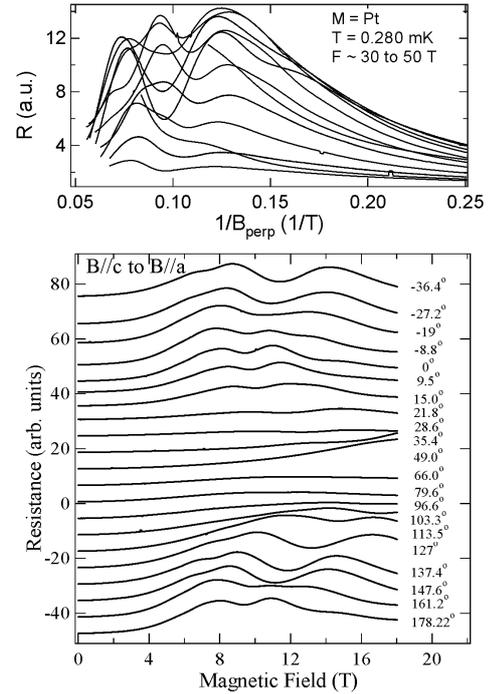

Figure 8. Angular dependence of low field structure in pressurized (5 kbar) $(Per)_2Pt(mnt)_2$ at 0.28 K. Upper panel: data presented vs. inverse perpendicular (to a-b plane) field.

Although there is no rigorous relationship, the data follow in an approximate 1/B periodicity with a frequency in the range 30 to 50 T, which corresponds to about 0.5% of the first Brillouin zone. The relative amplitudes of the three peaks observed in this quasi-Shubnikov-de Hass (SdH) behavior appear to vary with field orientation. The temperature dependence of the peaks is only weakly activated (see Ref. [16]). Under the assumption that these are indeed SdH oscillations associated with a very small pocket, we can propose a possible mechanism. If we consider the highly Q1D Fermi surface (FS) associated with the a-b plane, then it is unlikely that a pressure of 5 kbar could produce sufficient warping to close the Q1D sheets at the Brillouin zone edge since the "momentum gap" is $2k_F$. However, Canadell et al.[3] have recently considered the effects of the interaction between the four perylene chains in the unit cell, which produce a zone folding effect at $\pm k_F$, shown schematically in Fig. 9.

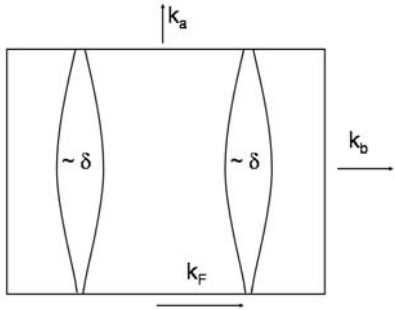

Figure 9. Simplified Fermi surface topology for $(Per)_2Pt(mnt)_2$ in the a-b plane (after Ref. [3].)

In this picture, the momentum gap between the zone-folded Fermi surface sheets is of order $\delta \sim 0.002$ Å$^{-1}$. Hence for modest pressures, if the two FS sheets meet, then closed Q2D orbits could appear. An estimate of the extremal area is $\delta \, 2\pi/a$, which is about 0.5 % of the first BZ, in accord with the observed SdH frequency. The angular dependence shown in Fig. 8 is also consistent with a Q2D closed orbit. It is interesting to speculate about these effects in both materials at higher pressures, and also their behavior in a field approaching 50 T, since this would put these pressurized orbital systems in the quantum limit. Further work is underway to explore this possibility.

**Summary**
The general effect of high magnetic fields on the CDW ground states of the $Per_2($Au, Pt$)(mnt)_2$ compounds is to suppress $T_{MI}$ nearly to zero at a field corresponding to the Pauli field. For the non-magnetic M=Au compound where the field is parallel to the conducting chains (B//b), only a residual marginally activated state remains in the range above $B_P$. However, for a finite field component perpendicular to the b-axis, orbital behavior appears, and the magnetoresistance becomes strongly activated and the resistance increases monotonically to the highest fields studied so far (60T). In the case for M=Pt, a second activated phase above $B_P$ appears for all field orientations we have so far studied. This phase, termed a field induced charge density wave, FICDW, also shows orbital behavior in tilted fields and in other measurements[16], but does not originate only from orbital mechanisms. The intermediate phase between the CDW and FICDW phases shows metallic character for B//c in a narrow range of field around $B_P$, and activated for other field directions. Comparing M = Au and Pt, the presence of magnetic chains may cause the more complex FICDW behavior, but at present the magnetic field effects on the spin-Peierls chains, and their interaction with the perylene chains is not yet known. Recalling that the Peierls transition is the dominant mechanism, and drives the spin-Peierls transition, the Pt chains may only be a perturbation on the more general nature of the high field phases. For M = Pt, no distinct features in the magnetization have yet been associated with spin-flop or spin flip type transitions. At high fields, the torque magnetization indicates that thermodynamic changes occur at the primary boundaries in the CDW and FICDW phases for M = Pt. (We have not yet made this measurement in the Au compound.) Due to the nature of the torque signal in a uniform field, these features are expected to arise from moments that are not aligned with the field. Hence they may arise from either orbital effects or spin configurations along certain directions. Although the presence of orbital effects seems unusual for a highly 1D system, the band structure is predicted to have interchain bandwidth that may become important at low temperatures. Due to the different interaction between the four perylene chains in the unit cell, the zone folding of the Fermi surface along the b-axis will cause further deviations from strictly 1D behavior. Although more systematic pressure studies are in progress to explore a more systematic increase in imperfect nesting parameters, data above 5 kbar suggest the possibility of stabilizing closed 2D orbits. If confirmed, this would give further evidence that the band calculations[3] do indeed describe the perylene compounds, and this would also show that at ambient pressure, the interchain band width plays a significant role, potentially in the orbital effects evident in the tilted field dependence of the phase diagrams shown in Fig. 2.

|  | Au | Cu | Pt | Pd |
|---|---|---|---|---|
| a (Å); $t_a$ (meV) | 16.602 | 17.5 | 16.612; 2 | 16.469 |
| b (Å); $t_b$ (meV) | 4.194 | 4.17 | 4.1891; 150 | 4.1891 |
| c (Å); $t_c$ (meV) | 25.546 | 25.5 | 26.583; 0 | 26.64 0 |
| Spin | 0 | 0 | 1/2 | 1/2 |
| $\sigma_{RT}$ (S/cm) | 700 | 700 | 700 | 300 |
| $S_{RT}$ (μV/K) | 32 | 38 | 32 | 32 |
| $T_{MI}$ (K) | 12.2 | 32 | 8.2 | 28 |
| $2\Delta$ (meV) | 3.5 | 20 | 8.6 | − |
| $10^4 \chi_{P-RT}$ (emu/mol) | 1.8 | 1.8 | 15.5 | 9.0 |
| $J/k_B$ (K) | − | − | 15 | 75 |
| $E_{th}$ (V/cm) | ~0.5 | − | ~15 | − |
| q (b*=2π/b) | − | b*/4 | b*/2 | b*/2 |
| $\gamma_{MF}$ | 0.134 | 0.21 | − | − |
| $B_P(T)$−BCS | 22.5 | 58.9 | 15.1 | 51.5 |
| $B_P(T)*$ | ~33 T | − | ~20 T | − |

Table 1. Representative parameters for selected Per$_2$M(mnt)$_2$ compounds. (After [2, 3] and references therein.) * B$_P$ taken from resistance minimum between CDW and FICDW phases in Refs. [14-16].

**Acknowledgements**

Supported by NSF-DMR 02-03532 and a NSF GK-12 Fellowship (DG). NHMFL is supported by the NSF and the State of Florida. Work in Portugal was supported by FCT under contract POCT/FAT/39115/2001.

**References**
[1] L. Alcacer and A. H. Maki, J. Phys. Chem. **78**, (1974)215 - 217.
[2] M. Almeida and R. T. Henriques, in *Handbook of Organic Conductive Molecules and Polymers*, ed. H. Nalwa (Wiley, New York, 1997) pp. 87-149.
[3] E. Canadell, M. Almeida, and J. S. Brooks, Eur. Phys. J. B **42**, (2004)R453.
[4] T. Ishiguro, K. Yamaji, and G. Saito, *Organic Superconductors II* (Springer-Verlag, Berlin, Heidelberg, New York, 1998).
[5] V. Gama, R. T. Henriques, M. Almeida, and J. P. Pouget, Synth. Met. **55-57**, (1993)1677.
[6] J. C. Xavier, R. G. Pereira, E. Miranda, and I. Affleck, Phys. Rev. Lett. **90**, (2003)247204.
[7] E. B. Lopes, M. J. Matos, R. T. Henriques, M. Almeida, and J. Dumas, Eur. Phys. Letters **27**, (1994)241.
[8] E. B. Lopes, M. J. Matos, R. T. Henriques, M. Almeida, and J. Dumas, Synth. Met. **70**, (1995)1267.
[9] E. B. Lopes, M. J. Matos, R. T. Henriques, M. Almeida, and J. Dumas, J. Phys. I **6**, (1996)2141.
[10] G. Bonfait, E. B. Lopes, M. J. Matos, R. T. Henriques, and M. Almeida, Solid State Commun. **80**, (1991)391.
[11] M. Matos, G. Bonfait, R. T. Henriques, and M. Almeida, Phys. Rev. B **54**, (1996)15307.
[12] W. Dieterich and P. Fulde, Z. Physik **265**, (1973)239
[13] D. C. Johnston, Phys. Rev. Lett. **52**, (1984)2049
[14] D. Graf, J. S. Brooks, E. S. Choi, S. Uji, J. C. Dias, M. Almeida, and M. Matos, Phys. Rev. B **69**, (2004)125113.
[15] D. Graf, E. S. Choi, J. S. Brooks, M. Matos, R. T. Henriques, and M. Almeida, Phys. Rev. Lett. **93**, (2004)076406.
[16] J. S. Brooks, D. Graf, E. S. Choi, M. Almeida, J. C. Dias, R. T. Henriques, and M. Matos, Curr. Appl. Phys., in press, and cond-mat/0501735, (2005)
[17] M. Fujita, K. Machida, and H. Nakanishi, J. Phys. Soc. Japan **54**, (1985)3820.
[18] D. Zanchi, A. Bjelis, and G. Montambaux, Phys. Rev. B **53**, (1996)1240.
[19] M. Kartsovnik, D. Andres, P. Grigoriev, W. Biberacher, and H. Müller, Physica B: Condensed Matter **346-347**, (2003)368.
[20] M. V. Kartsovnik, D. Andres, W. Biberacher, P. D. Grigoriev, E. A. Shuberth, and H. Muller, J. Phys. IV France **114**, (2004)191.
[21] D. Andres, M. V. Kartsovnik, P. D. Grigoriev, W. Biberacher, and H. Muller, Phys. Rev. B **68**, (2003)291191.
[22] D. Andres, M. V. Kartsovnik, W. Biberacher, H. Weiss, E. Balthes, H. Müller, and N. Kushch, Phys. Rev. B **64**, (2001)161104.
[23] R. H. McKenzie, cond-mat/9706235, (1997)
[24] A. G. Lebed, JETP Lett. **78**, (2003)138.
[25] R. T. Henriques, Ph.D. Thesis **IST, Lisbon**, (1986)
[26] H. Tanaka, M. Tokumoto, S. Ishibashi, D. Graf, E. S. Choi, J. S. Brooks, S. Yasuzuka, Y. Okano, H. Kobayashi, and A. Kobayashi, J. Am. Chem. Soc. **126**, (2004)10518.
[27] N.Mitsu, K. Yamaya, M. Almeida, and R. T. Henriques, J.Phys.Chem.Solids **in press**, (2005).